\input{aipcheck}

\documentclass[
    ,final            
  ]
  {aipproc}

\layoutstyle{6x9}

\begin{document}

\title{The $f_0(1370)$, $f_0(1710)$, $f_2(1270)$, $f_2'(1525)$, and $K_2^*(1430)$ as
dynamically generated states from vector meson - vector meson interaction}

\classification{13.75.Lb, 12.40.Yx, 13.20.Gd, 13.20.-v }
\keywords      {coupled-channel unitarity, vector meson--vector meson interaction, radiative decays, $J/\psi$ decays}

\author{L. S. Geng}{
  address={School of Physics and Nuclear Energy Engineering, Beihang University,  Beijing 100191,  China\\
Physik Department, Technische Universit\" at M\"unchen, D-85747 Garching, Germany}
}

\author{E. Oset}{
  address={Departamento de F\'{\i}sica Te\'orica and IFIC, Universidad de
Valencia-CSIC, E-46071 Spain}}

\author{R. Molina}{
  address={Departamento de F\'{\i}sica Te\'orica and IFIC, Universidad de
Valencia-CSIC, E-46071 Spain}}

\author{A. Mart\'inez Torres}{
  address={Departamento de F\'{\i}sica Te\'orica and IFIC, Universidad de
Valencia-CSIC, E-46071 Spain}}

\author{T. Branz}{
  address={Institut f\"ur Theoretische Physik,
Universit\"at T\"ubingen, Auf der Morgenstelle 14, D-72076 T\"ubingen, Germany}}

\author{F. K. Guo}{
  address={Institut f\"{u}r Kernphysik and J\"ulich Center
             for Hadron Physics,\\ Forschungszentrum J\"{u}lich,
             D--52425 J\"{u}lich, Germany}}
\author{L. R. Dai}{
  address={ Department of Physics, Liaoning Normal University, Dalian, 116029, China}}

  \author{B. X. Sun}{
  address={Institute of Theoretical Physics, College of Applied Sciences,
Beijing University of Technology, Beijing 100124, China.}}

\begin{abstract}
 We report on some recent developments in understanding the
nature of the low-lying mesonic resonances $f_0(1370)$, $f_0(1710)$, $f_2(1270)$,
$f_2'(1525)$, and $K_2^*(1430)$. In particular
we show that these five resonances can be dynamically generated from vector meson--vector meson interaction
 in a coupled-channel unitary approach, which utilizes the phenomenologically
very successful hidden-gauge Lagrangians to produce the interaction kernel between two vector mesons, which is
then unitarized by the Bethe-Salpeter-equation method. The data on the strong decay branching ratios, total decay widths, and radiative decay widths of these
five states,
and on related $J/\psi$ decay processes can all be well described by such an approach. We also make predictions,
compare them with the results of earlier studies, and highlight observables that if measured can be used
to distinguish different pictures of these resonances.
\end{abstract}

\maketitle


\section{Introduction}
In recent years, coupled-channel unitarity
has been shown to play a very important role in understanding
final state interactions and the nature of certain resonances which
do not fit easily into naive valence quark models, where mesons consist of a pair of
quark-antiquark and baryons three quarks.
By combining coupled-channel unitarity with chiral perturbation theory,
 the so-called unitary chiral theories (U$\chi$PT) have proven to be a very powerful tool to describe some of such resonances, e.g., the $f_0(600)$~\cite{Oller:1997ti}
 and the $\Lambda(1405)$~\cite{Kaiser:1995eg}. The coupled-channel unitary methods can also be extended to employ interaction kernels provided
 by other Langrangians, e.g., the hidden-gauge Langrangians~\cite{Molina:2008jw,Geng:2008gx,Oset:2009vf,Sarkar:2009kx} and Lagrangians inspired by SU(4) flavor symmetry~\cite{Molina:2009eb,Gamermann:2006nm} or SU(6)[SU(8)] spin-flavor
 symmetry~\cite{GarciaRecio:2008dp,GarciaRecio:2010ki}.

 An interesting development in this line of research is the description of the interactions of the
 lowest-lying vector mesons either with themselves~\cite{Molina:2008jw,Geng:2008gx} or with other hadrons, e.g., the lowest-lying
 octet~\cite{Oset:2009vf} and decuplet~\cite{Sarkar:2009kx} baryons. It was shown in Ref.~\cite{Geng:2008gx} that the interaction between two vector mesons naturally generates 11
 resonances, which include not only the two scalar resonances $f_0(1370)$ and
 $f_0(1710)$, whose nature are still hotly debated both theoretically and experimentally, but also the tensor resonances
 $f_2(1270)$, $f_2'(1525)$, and $K_2^*(1430)$, which for a long time have been firmly believed to
 be simple $q\bar{q}$ states. Whether such a dynamical picture of these states is correct or partially correct can
 only be checked by studying as extensively as possible the corresponding theoretical implications and comparing them with data and those
 predicted by other approaches, such as naive quark models.
For this purpose we have studied not only their strong-decay branching ratios~\cite{Geng:2008gx} but also their radiative decay widths~\cite{Branz:2009cv}. Furthermore
 we have looked at the abundant $J/\psi$ decay data where $J/\psi$ decays into one of these five resonances with either
a vector meson~\cite{MartinezTorres:2009uk} or a photon~\cite{Geng:2009iw}.

This paper is organized as follows. In section 2, we briefly summarize the coupled-channel unitary approach, within which the $f_0(1370)$, $f_0(1710)$, $f_2(1270)$, $f_2'(1525)$,  and $K_2^*(1430)$ are dynamically generated. In Section
3, we compare their strong decay branching ratios with available data. In Section 4, we discuss the related
$J/\psi$ decays. In Section 5, we calculate the radiative decay widths and compare with data and the predictions of other approaches. A brief summary is given in Section 6.

\section{Theoretical Framework}
For the sake of completeness, in this section we provide a short description of the coupled-channel unitary approach, in which the $f_0(1370)$, $f_0(1710)$, $f_2(1270)$, $f_2'(1525)$,  and $K_2^*(1430)$ are dynamically
generated. For more details, we refer the readers to Ref.~\cite{Geng:2008gx}. The starting point of the coupled-channel unitary approach is  the interaction kernel. In the present case this is provided by the hidden-gauge Lagrangians~\cite{Bando:1984ej},
which are known to describe very well the interactions between vector mesons and other hadrons including the vector mesons themselves. Once a suitable interaction kernel is chosen, the next step is to unitarize it. Here one can choose from a number of different procedures, such as the Bethe-Salpeter-equation method, which we chose, the N/D method, and the inverse amplitude method (IAM). For a short discussion of the similarities and differences between
these three unitarization procedures, see, e.g., Ref.~\cite{Geng:2008ag}

The parameters of the coupled-channel unitary approach are related to
the regularization of the loops appearing in the unitarization procedure, which are cutoff values in the cutoff regularization
 method or subtraction constants in the dimensional regularization method. In the case of a light meson (such as the $\pi$) interacting with a heavy particle (such as the nucleon and the $D$ mesons), model-independent determination of these parameters has been claimed. On the other hand, in our present case we simply treat them as free parameters with their values constrained by the observation that
if one uses the cutoff method to regularize the loop integrals, the value of the cutoff should be around 1 GeV, to be consistent with what one would expect from a phenomenological point of view.

As explained in detail in Ref.~\cite{Geng:2008gx}, this is exactly the procedure we followed. That is to say, we first performed the calculation using a cutoff value, $\Lambda$, of about 1 GeV and allowed $\Lambda$ to vary within a reasonable range, e.g., 10$\sim$20$\%$. This way we found that 11 resonances got dynamically generated in 9 isospin-strangeness channels and some of them can be easily associated to the experimentally well-known states, such as the $f_0(1370)$, $f_0(1710)$, $f_2(1270)$, $f_2'(1525)$, and $K_2^*(1430)$. We then adopted the dimensional regularization method to regularize the loop integrals in order to be able to perform analytic continuation of the amplitudes to the complex plane. The values of the subtraction constants were fixed in such a way that the results obtained using the cutoff method were reproduced. We have slightly tuned the values of the subtraction constants to reproduce the masses of
the three tensor states. We must stress here that in our determination of the model parameters we have not fitted the branching ratios of these states and therefore they
are predictions fixed completely by the dynamics built into the model.

\section{The (strong) partial decay widths and branching ratios}
A detailed description of the calculation of the partial decay widths can
be found in Ref.~\cite{Geng:2009gb}. In Table 1, we tabulate the (strong) branching ratios of
the states $f_0(1370)$, $f_0(1710)$, $f_2(1270)$, $f_2'(1525)$, $K_2^*(1430)$ in comparison with
available data. It is clear that our results for the
two $f_2$ states agree very well with the data. For the $f_0(1370)$, according to the PDG~\cite{Amsler:2008zz},
the $\rho\rho$ mode is dominant.
In our approach, however, the $\pi\pi$ mode is dominant, which is consistent with the results of
Ref.~\cite{Albaladejo:2008qa}
and the recent analysis of D. V. Bugg~\cite{Bugg:2007ja}. For the $f_0(1710)$, using the branching
ratios given in Table 1, we obtained $\Gamma(\pi\pi)/\Gamma(K\bar{K})<1\%$ and $\Gamma(\eta\eta)/\Gamma(K\bar{K})\sim49\%$.
On the other hand, the PDG gives the following
averages: $\Gamma(\pi\pi)/ \Gamma(K\bar{K}) =0.41_{-0.17}^{+ 0.11}$,
and $\Gamma(\eta\eta)/\Gamma(K \bar{K})=0.48 \pm0.15$~\cite{Amsler:2008zz}.  Our calculated branching ratio
for the $\eta\eta$ channel is in agreement with their average, while the ratio for
the $\pi\pi$ channel is much smaller. However, we notice that
the above PDG $\Gamma(\pi\pi)/ \Gamma(K\bar{K})$ ratio is taken from the BES data on
$J/\psi\rightarrow \gamma\pi^+\pi^-$~\cite{Ablikim:2006db}, which comes from
a partial wave analysis that includes seven resonances. On the other hand,
the BES data on $J/\psi\rightarrow\omega K^+K^-$~\cite{Ablikim:2004st} give an upper limit
$\Gamma(\pi\pi)/ \Gamma(K\bar{K})<11\%$ at the $95\%$ confidence level. Clearly more analysis is
needed to settle the issue.

In Table 1, one can see that the dominant decay mode of the $K^*_2(1430)$ is $K\pi$ both theoretically and experimentally. However, other modes, such as $\rho K$, $K^*\pi$, and $K^*\pi\pi$,
account for half of its decay width according to the PDG~\cite{Amsler:2008zz}. This is consistent with the fact that our $K^*_2(1430)$ is narrower than its experimental counterpart~\cite{Geng:2008gx}.
\begin{table}
      \renewcommand{\arraystretch}{1.4}
     \setlength{\tabcolsep}{0.2cm}
\caption{Branching ratios of the $f_0(1710)$, $f_0(1370)$, $f_2(1270)$, $f'_2(1525)$, and $K^*_2(1430)$ in comparison with data~\cite{Amsler:2008zz}.}
\begin{tabular}{c|cc|cc|cc|cc}
\hline\hline
 &\multicolumn{2}{c|}{$\Gamma(\pi\pi)/\Gamma(\mathrm{total})$}&\multicolumn{2}{c|}{$\Gamma(\eta\eta)/\Gamma(\mathrm{total})$}&\multicolumn{2}{c|}{$\Gamma(K\bar{K})/\Gamma(\mathrm{total})$}
&\multicolumn{2}{c}{$\Gamma(\mathrm{VV})/\Gamma(\mathrm{total})$}\\
& Our model & Data & Our model & Data & Our model & Data & Our model & Data\\
\hline
$f_0(1370)$ & $\sim72\%$ & & $<1\%$ & & $\sim10\%$ & &$\sim18\%$ &\\
$f_0(1710)$ & $<1\%$    &  & $\sim27\%$ &  & $\sim55\%$&& $\sim18\%$ &\\\hline
$f_2(1270)$ & $\sim88\%$ & $84.8\%$ & $ <1\%$ & $<1\%$ & $\sim 10\%$&$4.6\%$ & $<1\%$ &\\
$f'_2(1525)$ &$<1\%$ & $0.8\%$ & $\sim21\%$ & $10.4\%$ & $\sim 66\%$& $88.7\%$ & $\sim13\%$&\\
$K^*_2(1430)$ & $\sim93\%$ & $49.9\%$  & $\sim5\%$ & $<1\%$ & $\sim2\%$ \\
 \hline\hline
\end{tabular}
\end{table}

\section{Related $J/\psi$ decays}
\subsection{$J/\psi$ decays into a vector meson $\rho^0$, $\omega$, $\phi$ and one of the tensor
mesons $f_2(1270)$, $f_2'(1525)$, and $K_2^*(1430)$.}

$J/\psi$ decays offer a good opportunity to test the dynamical picture
 of the $f_0(1370)$, $f_0(1710)$, $f_2(1270)$, $f_2'(1525)$, and
 $K_2^*(1430)$. One of such processes is $J/\psi$ decay into a vector meson $\rho^0$, $\omega$, or $\phi$ and
one of the tensor mesons $f_2(1270)$, $f_2'(1525)$, $K_2^*(1430)$. Furthermore, if we are only interested in
the ratios of the partial decay widths we could ignore absolute normalization and greatly simplify the calculation by noting that $J/\psi$ is a SU(3) singlet and the tensor resonances are
dynamically generated from vector meson--vector meson interactions. More details can be found in Ref.~\cite{MartinezTorres:2009uk}.

To study the ratios of different decay rates, e.g.,
\begin{equation}
R_{1}\equiv\frac{\Gamma_{J/\psi\to\phi f_{2}(1270)}}{\Gamma_{J/\psi\to\phi f^\prime_{2}(1525)}},\quad
R_{2}\equiv\frac{\Gamma_{J/\psi\to\omega f_{2}(1270)}}{\Gamma_{J/\psi\to\omega f^\prime_{2}(1525)}},\label{ratios1}
\end{equation}
\begin{equation}
R_{3}\equiv\frac{\Gamma_{J/\psi\to\omega f_{2}(1270)}}{\Gamma_{J/\psi\to\phi f_{2}(1270)}},\quad
R_{4}\equiv\frac{\Gamma_{J/\psi\to K^{*\,0} \bar{K}^{*\,0}_{2}(1430)}}{\Gamma_{J/\psi\to\omega f_{2}(1270)}},\label{ratios2}
\end{equation}
 we needed only one parameter $\nu$, which can be fixed by fitting our predictions to data,
Upon minimization of the $\chi^2$ function we obtained an optimal solution: $\nu=1.45$. For our estimate of the theoretical uncertainties, see Ref.~\cite{MartinezTorres:2009uk}.
\begin{table}[b]
\caption{Comparison between the experimental and the theoretical results.}
\begin{tabular}{ccc}
\hline\hline
&Experiment&Theory\\
\hline
\\
$R_{1}$&0.22 - 0.47 $(0.33^{+0.14}_{-0.11})$&0.13 - 0.61 $(0.28^{+0.33}_{-0.15})$\\[1.5ex]
$R_{2}$&12.33 - 49.00 $(21.50^{+27.50}_{-9.17})$&2.92 - 13.58 $(5.88^{+7.70}_{-2.96})$\\[1.5ex]
$R_{3}$&11.21 - 23.08 $(15.85^{+7.23}_{-4.65})$&6.18 - 19.15 $(10.63^{+8.52}_{-4.45})$\\[1.5ex]
$R_{4}$&0.55 - 0.89 $(0.70^{+0.19}_{-0.15})$&0.83 - 2.10 $(1.33^{+0.77}_{-0.50})$\\[1.5ex]
\hline
\end{tabular}
\label{res}
\end{table}
As can be seen from Table 2, the overall agreement of our results with the data is reasonable. We obtained four independent ratios with just one parameter. On the other hand this parameter
can be related to $\lambda_{\phi}$ of Ref.~\cite{ulfjose} as done in Ref.~\cite{palochiang} through
\begin{equation}
\lambda_{\phi}=\sqrt{2}\Bigg (\frac{\nu-1}{\nu+2}\Bigg)
\end{equation}
which provides a value of $\lambda_{\phi}=0.18$, very close to the one obtained
 in Refs.~\cite{ulfjose,palochiang}, $\lambda_{\phi}=0.13-0.20$. Although we have different physics than in
 Ref.~\cite{ulfjose,palochiang} since we have the production of pairs of vector mesons
 rather than pseudoscalar mesons, and we have also tensor states rather than scalars,
 it is gratifying to see that the value of $\lambda_{\phi}$, which is a measure
 of the subdominant, double OZI suppressed, mechanism in $J/\psi\to \phi V^\prime V^\prime$ (see Fig.~1b of Ref.~\cite{MartinezTorres:2009uk}), is a small number, comparable in
 size and sign to the one obtained in Refs.~\cite{ulfjose,palochiang}.

 The success in the description of the experimental data is by no means trivial
and can be traced back to the particular couplings of the resonance to the
$V^\prime V^\prime$ states. Note that the important couplings to $\rho\rho$ and
$K^*\bar{K}^*$ have the same relative sign for the $f_{2}(1270)$ and opposite relative sign for the $f^{\prime}_{2}(1525)$ (see Table I of
Ref.~\cite{Geng:2008gx}). This feature is essential to the success of the results.
 Should all the couplings have the same sign it would have been impossible to
 get any reasonable fit to the data (see Fig.~7 of Ref.~\cite{MartinezTorres:2009uk}).

\subsection{$J/\psi$ decays into a photon and one of the $f_2(1270)$, $f_2'(1525)$, $f_0(1370)$, and $f_0(1710)$}
\begin{figure}[tb]
\includegraphics[scale=0.6]{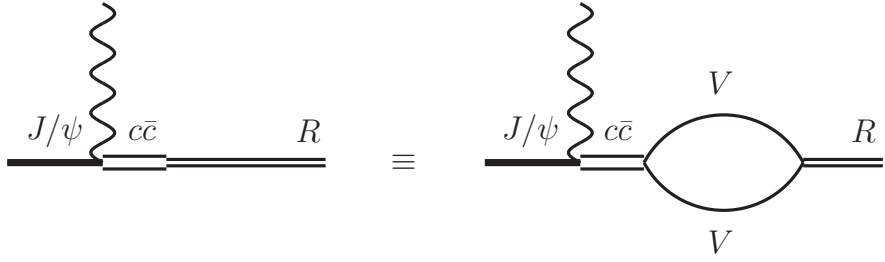}
\caption{Schematic representation of $J/\psi$ decay into a photon and one dynamically generated resonance.} \label{f4}
\end{figure}
In a similar way to what was shown above, one can also study the $J/\psi$ decays into a photon and one of the
$f_2(1270)$, $f_2'(1525)$, $f_0(1370)$, and $f_0(1710)$~\cite{Geng:2009iw}. Such a process is schematically shown in
Fig.1. To remove the dependence of the $c\bar{c}VV$ vertex on spin we constructed the following
two ratios
\begin{equation}
R_T=\Gamma_\mathrm{J/\psi\rightarrow \gamma f_2(1270)}/\Gamma_{J/\psi\rightarrow
\gamma f'_2(1525)},
\end{equation}
\begin{equation}
R_S=\Gamma_{J/\psi\rightarrow \gamma f_0(1370)}/\Gamma_{J/\psi\rightarrow\gamma
f_0(1710)}.
\end{equation}
The results are compared with available data in Table 3 with the theoretical uncertainties
estimated in Ref.~\cite{Geng:2009iw}. Our result for
$R_T$ is in reasonable agreement with data while there is no data for $R_S$. To show the differences between our dynamical
picture and the naive quark model where the $f_0(1370)$ and $f_0(1710)$ belong to the same scalar nonet while
the $f_2(1270)$ and $f_2'(1525)$ belong to the same tensor nonet, we have calculated $R_T$ and $R_S$ using
a rather naive quark model. The results obtained are also shown in Table 3. It is quite amazing to notice
that although the quark model yields a $R_T$ consistent with our prediction, it gives a quite different
$R_S$ than ours. Experimental measurement of $R_S$ will be
very useful to distinguish between these two different pictures of the $f_0(1370)$ and $f_0(1710)$.

\begin{table}[t]
      \renewcommand{\arraystretch}{1.5}
     \setlength{\tabcolsep}{0.3cm}
\caption{Ratios of  $R_T=\Gamma_\mathrm{J/\psi\rightarrow \gamma f_2(1270)}/\Gamma_{J/\psi\rightarrow
\gamma f'_2(1525)}$ and $R_S=\Gamma_{J/\psi\rightarrow \gamma f_0(1370)}/\Gamma_{J/\psi\rightarrow\gamma
f_0(1710)}$ within the molecular model and the quark model
 in comparison with data~\cite{Amsler:2008zzb}.
\label{table:ratio}}
\begin{tabular}{c|ccc}\hline\hline
 & Molecular picture & Quark model & Data\\\hline
$R_T$ & $2\pm 1$ & $2.2$ & $3.18^{+0.58}_{-0.64}$ \\
$R_S$ & $1.2\pm 0.3$ & $2.2-2.5$ & \\
$R_S/R_T$ & $0.6\pm 0.1$ & $1-1.1$ & \\
 \hline\hline
    \end{tabular}
\end{table}

\section{Radiative decays of the $f_0(1370)$, $f_0(1710)$, $f_2(1270)$, $f_2'(1525)$, and $K_2^*(1430)$}
\begin{figure}[t]
\centerline{\includegraphics[scale=0.6]{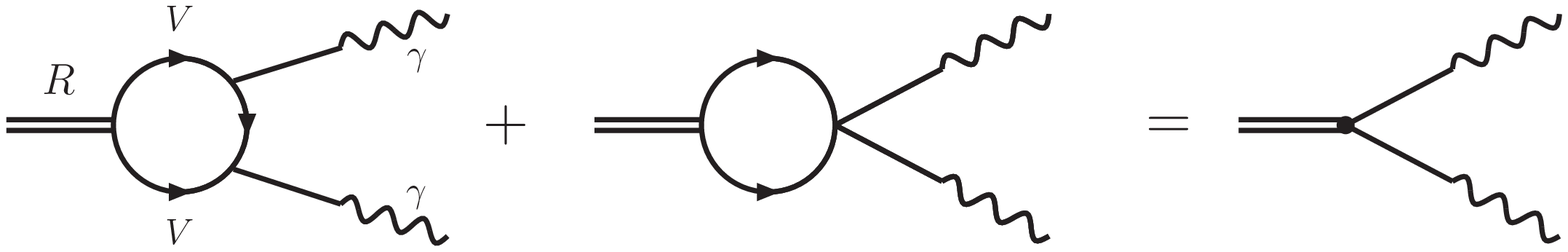}}
\caption{Two-photon decay of a dynamically generated resonance from vector meson-vector meson interaction.\label{fig:twogamma}}
\end{figure}
Radiative decay provides a very clean probe of the structure of hadronic states. For instance, the non-observation of
the $f_0(1500)$ decaying into two photons has been used to support
its dominant glue nature~\cite{Amsler:2002ey}. In Ref.~\cite{Branz:2009cv}, we have calculated
the two-photon ($\gamma\gamma$) and the vector-meson--photon ($V\gamma$) decay widths of the 11 dynamically
generated states from vector meson--vector meson interactions. Because these 11
states are built from vector meson --vector meson interactions and because in the hidden-gauge Lagrangians
photons do not couple directly to charged vector mesons but must first convert into neutral
vector mesons, the calculation of the radiative decay widths is quite straightforward. As shown in Fig.~2, one simply
needs the couplings of the resonance of interest to the vector meson -- vector meson coupled channels. This has
been done in Ref.~\cite{Geng:2008gx}. With these couplings,
one can easily obtain the $\gamma\gamma$ and $V\gamma$ decay widths as
\begin{eqnarray}
\Gamma_{\gamma\gamma}&=&\frac{1}{2S+1}\frac{1}{16\pi M_R}\frac{1}{2}\times
\sum\limits_{\mbox{polarization}}|T^{(R)}_{\gamma\gamma}|^2,\\
  \Gamma_{V\gamma }&=&\frac{1}{2S+1}\frac{1}{8\pi M_R}\frac{|p_\gamma|}{M_R}\times
\sum\limits_{\mbox{polarization}}|T^{(R)}_{V\gamma}|^2,
\end{eqnarray}
where $M_R$ is the resonance mass, $p_\gamma$ is the photon momentum in
the rest frame of the resonance $R$, $S$ is the resonance spin,  and $T^{(R)}_{\gamma\gamma}$ ($T^{(R)}_{V\gamma}$) is defined
in Ref.~\cite{Branz:2009cv}.

Our predictions for the $VV$ and $V\gamma$ decay widths of the $f_0(1370)$, $f_0(1710)$, $f_2(1270)$,
$f_2'(1525)$, and $K_2^*(1430)$ are tabulated in Tables I-VIII of Ref.~\cite{Branz:2009cv}.
Our results agree reasonably well with the data, but show
distinct behavior compared to the predictions of other approaches. For instance, the $f_2'(1525)\rightarrow \rho^0\gamma$ and
$f_2'(1525)\rightarrow\omega\gamma$ partial decay widths (see Table 4) are quite different from those predicted by the covariant oscillator quark model (COQM). An
experimental measurement of these two decay modes, particularly their ratio, should be very useful to distinguish the
two different pictures of the $f_2'(1525)$. Similar distinct results have been observed for the
$f_0(1370)$, $f_0(1710)$, $f_2(1270)$, and $f_2'(1525)$~\cite{Branz:2009cv}.

\begin{table}[t]
      \renewcommand{\arraystretch}{1.6}
     \setlength{\tabcolsep}{0.3cm}
     \centering
     \caption{Radiative decay widths of the $f'_2(1525)$ obtained
     in the present work in comparison with those obtained in the covariant oscillator quark model (COQM)~\cite{Ishida:1988uw}.\label{table:f2pcomparison}}
         \begin{tabular}{cccc}
     \hline\hline
&COQM~\cite{Ishida:1988uw} &Present
work\\\hline
$f'_2(1525)\to\gamma \gamma$& &0.05\\
$f'_2(1525)\to\rho^0 \gamma$&4.8 &72\\
$f'_2(1525)\to \omega \gamma$&0&224\\
$f'_2(1525)\to\phi\gamma$ &104& 286\\
\hline\hline
    \end{tabular} 
\end{table}

\section{Summary}
The nature of a hadronic state is definitely more complex than what one used to assume.
In a naive quark model, baryons consist of three quarks ($qqq$) and mesons a pair of quark and anti-quark ($q\bar{q}$).
In recent years, studies have found evidences that baryons and mesons often contain multi-quark components, which in certain cases
may even be more important than the $qqq$ and $q\bar{q}$ structures. In the $u$, $d$, $s$ flavor sector,
the $f_0(600)$ and the $\Lambda(1405)$ are believed to be such states. The newly experimentally measured $X$, $Y$, $Z$ particles are examples
in the charmonium sector.

Coupled-channel unitary approaches are particularly suitable for describing $s$-wave resonances whose
wave function contains large meson-meson or meson-baryon components. In the past few years,
many resonances have been generated in coupled-channel unitary approaches and they often are referred to as
"dynamically generated resonances." However, generating resonances is only
the first step. One has to test as extensively as possible the consequences of such descriptions. In this talk, we have
reported on a series of recent works exploring the possible dynamical picture of the $f_0(1370)$, $f_0(1710)$, $f_2(1270)$,
$f_2'(1525)$, and $K_2^*(1430)$. Our studies have shown that all existing data are consistent with the dynamical picture and
therefore the wave function of these five states may contain important multi-quark components in the form
of vector meson-vector meson configuration. Furthermore, for certain observables our studies have shown that this dynamical picture predicts distinct and (in principle)
detectable patterns compared to, e.g., various quark models.
Experimental verification of these predictions will be very useful to distinguish between different pictures
of the $f_0(1370)$, $f_0(1710)$, $f_2(1270)$,
$f_2'(1525)$, and $K_2^*(1430)$.

\begin{theacknowledgments}
L. S. Geng acknowledges support from the Alexander von Humboldt foundation (Germany) and the Fundamental Research Funds for the Central Universities (China).
\end{theacknowledgments}


\bibliographystyle{aipproc}   

\begin{thebibliography}{99}

\bibitem{Oller:1997ti}
  J.~A.~Oller and E.~Oset,
  Nucl.\ Phys.\  A {\bf 620}, 438 (1997)
  [Erratum-ibid.\  A {\bf 652}, 407 (1999)];
  N.~Kaiser,
  Eur.\ Phys.\ J.\  A {\bf 3}, 307 (1998);
  V.~E.~Markushin,
  Eur.\ Phys.\ J.\  A {\bf 8}, 389 (2000);
  A.~Dobado and J.~R.~Pelaez,
  Phys.\ Rev.\  D {\bf 56}, 3057 (1997);
  J.~A.~Oller, E.~Oset and J.~R.~Pelaez,
  Phys.\ Rev.\  D {\bf 59}, 074001 (1999)
  [Erratum-ibid.\  D {\bf 60}, 099906 (1999);  Erratum-ibid. D75, 099903 (2007)].

\bibitem{Kaiser:1995eg}
  N.~Kaiser, P.~B.~Siegel and W.~Weise,
  Nucl.\ Phys.\  A {\bf 594}, 325 (1995);
  E.~Oset and A.~Ramos,
  Nucl.\ Phys.\  A {\bf 635}, 99 (1998);
  J.~A.~Oller and U.~G.~Meissner,
  Phys.\ Lett.\  B {\bf 500}, 263 (2001);
  C.~Garcia-Recio, M.~F.~M.~Lutz and J.~Nieves,
  Phys.\ Lett.\  B {\bf 582}, 49 (2004);
  D.~Jido, J.~A.~Oller, E.~Oset, A.~Ramos and U.~G.~Meissner,
  Nucl.\ Phys.\  A {\bf 725}, 181 (2003);
  C.~Garcia-Recio, J.~Nieves and L.~L.~Salcedo,
  Phys.\ Rev.\  D {\bf 74}, 034025 (2006);
  T.~Hyodo, S.~I.~Nam, D.~Jido and A.~Hosaka,
  Phys.\ Rev.\  C {\bf 68}, 018201 (2003).


\bibitem{Molina:2008jw}
 R.~Molina, D.~Nicmorus and E.~Oset,
 Phys.\ Rev.\  D {\bf 78}, 114018 (2008).

\bibitem{Geng:2008gx}
  L.~S.~Geng and E.~Oset,
  Phys.\ Rev.\  D {\bf 79}, 074009 (2009).

\bibitem{Oset:2009vf}
  E.~Oset and A.~Ramos,
  Eur.\ Phys.\ J.\  A {\bf 44}, 445 (2010).




\bibitem{Sarkar:2009kx}
  S.~Sarkar, B.~X.~Sun, E.~Oset and M.~J.~V.~Vacas,
  Eur.\ Phys.\ J.\  A {\bf 44}, 431 (2010).


\bibitem{Molina:2009eb}
  R.~Molina, H.~Nagahiro, A.~Hosaka and E.~Oset,
  Phys.\ Rev.\  D {\bf 80}, 014025 (2009).

\bibitem{Gamermann:2006nm}
  D.~Gamermann, E.~Oset, D.~Strottman and M.~J.~Vicente Vacas,
  Phys.\ Rev.\  D {\bf 76}, 074016 (2007).


\bibitem{GarciaRecio:2008dp}
  C.~Garcia-Recio, V.~K.~Magas, T.~Mizutani, J.~Nieves, A.~Ramos, L.~L.~Salcedo and L.~Tolos,
  Phys.\ Rev.\  D {\bf 79}, 054004 (2009).


\bibitem{GarciaRecio:2010ki}
  C.~Garcia-Recio, L.~S.~Geng, J.~Nieves and L.~L.~Salcedo,
  arXiv:1005.0956 [hep-ph].

\bibitem{Branz:2009cv}
  T.~Branz, L.~S.~Geng and E.~Oset,
  Phys.\ Rev.\  D {\bf 81}, 054037 (2010).

\bibitem{MartinezTorres:2009uk}
  A.~Martinez Torres, L.~S.~Geng, L.~R.~Dai, B.~X.~Sun, E.~Oset and B.~S.~Zou,
  Phys.\ Lett.\  B {\bf 680}, 310 (2009).

\bibitem{Geng:2009iw}
  L.~S.~Geng, F.~K.~Guo, C.~Hanhart, R.~Molina, E.~Oset and B.~S.~Zou,
  Eur.\ Phys.\ J.\  A {\bf 44}, 305 (2010).


\bibitem{Bando:1984ej}
  M.~Bando, T.~Kugo, S.~Uehara, K.~Yamawaki and T.~Yanagida,
  Phys.\ Rev.\ Lett.\  {\bf 54}, 1215 (1985);
  M.~Bando, T.~Kugo and K.~Yamawaki,
  Phys.\ Rept.\  {\bf 164}, 217 (1988).

\bibitem{Geng:2008ag}
  L.~S.~Geng, E.~Oset, J.~R.~Pelaez and L.~Roca,
  Eur.\ Phys.\ J.\  A {\bf 39}, 81 (2009).

\bibitem{Geng:2009gb}
  L.~S.~Geng, E.~Oset, R.~Molina and D.~Nicmorus,
  PoS E {\bf FT09}, 040 (2009).

\bibitem{Amsler:2008zz}
  C.~Amsler {\it et al.}  [Particle Data Group],
  Phys.\ Lett.\  B {\bf 667}, 1 (2008).


\bibitem{Albaladejo:2008qa}
  M.~Albaladejo and J.~A.~Oller,
  Phys.\ Rev.\ Lett.\  {\bf 101}, 252002 (2008).


\bibitem{Bugg:2007ja}
  D.~V.~Bugg,
  Eur.\ Phys.\ J.\  C {\bf 52}, 55 (2007).

\bibitem{Ablikim:2006db}
  M.~Ablikim {\it et al.},
  Phys.\ Lett.\  B {\bf 642}, 441 (2006).


\bibitem{Ablikim:2004st}
  M.~Ablikim {\it et al.}  [BES Collaboration],
  Phys.\ Lett.\  B {\bf 603} (2004) 138.

\bibitem{ulfjose}
 U.~G.~Meissner and J.~A.~Oller,
 Nucl.\ Phys.\  A {\bf 679}, 671 (2001).

\bibitem{palochiang}
 L.~Roca, J.~E.~Palomar, E.~Oset and H.~C.~Chiang,
 Nucl.\ Phys.\  A {\bf 744}, 127 (2004).


\bibitem{Amsler:2008zzb}
  C.~Amsler {\it et al.}  [Particle Data Group],
  Phys.\ Lett.\  B {\bf 667}, 1 (2008).

\bibitem{Amsler:2002ey}
  C.~Amsler,
  Phys.\ Lett.\  B {\bf 541}, 22 (2002).


\bibitem{Ishida:1988uw}
  S.~Ishida, K.~Yamada and M.~Oda,
  Phys.\ Rev.\  D {\bf 40}, 1497 (1989).

\end{thebibliography}

\end{document}